\documentstyle[rotating,a4wide,epsfig,twoside,12pt]{article}
\def\vec#1{\mathchoice{\mbox{\boldmath$\mathrm\displaystyle#1$}}
{\mbox{\boldmath$\mathrm\textstyle#1$}}
{\mbox{\boldmath$\mathrm\scriptstyle#1$}}
{\mbox{\boldmath$\mathrm\scriptscriptstyle#1$}}}

\newcommand{\be}{\begin{equation}}
\newcommand{\ee}{\end{equation}}
\newcommand{\ba}{\begin{array}}
\newcommand{\ea}{\end{array}}
\newcommand{\bea}{\begin{eqnarray}}
\newcommand{\eea}{\end{eqnarray}}

\newsavebox{\TRS}
\sbox{\TRS}{\hspace{.5em} = \hspace{-1.8em}
                 \raisebox{1ex}{\mbox{\scriptsize TRS}} }

\newsavebox{\defgleich}
\sbox{\defgleich}{\ :=\ }

\newsavebox{\LSIM}
\sbox{\LSIM}{\raisebox{-1ex}{$\ \stackrel{\textstyle<}{\sim}\ $}}

\newsavebox{\GSIM}
\sbox{\GSIM}{\raisebox{-1ex}{$\ \stackrel{\textstyle>}{\sim}\ $}}

\newcommand{\lk}{\left}
\newcommand{\rk}{\right}
\newcounter{saveeqn}

\newcommand{\sssty}{\scriptscriptstyle}
\newcommand{\Del}{\Delta}

\newcommand{\sig}{\sigma}

\newcommand{\hpv}{\mbox{$H_{\sssty\mathrm PV}$}}
\newcommand{\hpve}{\mbox{$H_{\sssty\mathrm PV}^{\sssty (1)}$}}
\newcommand{\hpvz}{\mbox{$H_{\sssty\mathrm PV}^{\sssty (2)}$}}
\newcommand{\hpvez}{\mbox{$H_{\sssty\mathrm PV}^{\sssty (1,2)}$}}

\newcommand{\qwe}{\mbox{$Q_{\sssty W}^{\sssty (1)}$}}
\newcommand{\qwz}{\mbox{$Q_{\sssty W}^{\sssty (2)}$}}

\newcommand{\qwi}{\mbox{$Q_{\sssty W}^{\sssty (i)}$}}

\newcommand{\theW}{\theta_{\sssty W}}

\newcommand{\cE}{{\cal E}}

\newcommand{\cO}{{\cal O}}

\newcommand{\cS}{{\cal S}}

\newcommand{\deli}{\delta_i}

\newcommand{\dele}{\delta_1}
\newcommand{\delz}{\delta_2}
\newcommand{\delez}{\delta_{1,2}}

\newcommand{\unH}{\mbox{$\underline{\cal H}$}}

\newcommand{\eve}{\vec{e}_1}
\newcommand{\evz}{\vec{e}_2}
\newcommand{\evd}{\vec{e}_3}

\newcommand{\cEv}{\vec{\cal E}}

\newcommand{\cEvR}{\vec{\cal E}_{\mbox{\tiny R}}}

\newcommand{\etav}{\vec{\eta}}

\newcommand{\Rv}{\vec{R}}

\newcommand{\iKsig}{{\sssty (\sigma )}}

\newcommand{\iKi}{{\sssty (i)}}

\newcommand{\iKn}{{\sssty (0)}}
\newcommand{\iKe}{{\sssty (1)}}
\newcommand{\iKz}{{\sssty (2)}}
\newcommand{\iKd}{{\sssty (3)}}
\newcommand{\iKvar}[1]{{\sssty (#1)}}

\newcommand{\eeH}{\left.^1_1\mbox{H}\right.}

\date{}
\begin{document}
{\sloppy
%
%
%
\begin{titlepage}

\title{
{\normalsize 
\hfill HD--THEP--97--52
}
\vspace{1cm}
\\
{\LARGE\bf\sf New Observables\\
for Parity Violation in Atoms:\\
Energy Shifts in External Electric Fields\thanks{
Work supported by Deutsche Forschungsgemeinschaft, Project No.\ Na 296/1--1} \stepcounter{footnote}}
}

\author{
{\sc 
D. Bru\ss,\thanks{Present address: ISI, Villa Gualino, Viale Settimio Severo 65, 10133 Torino, Italy.}\ \
T. Gasenzer,\thanks{Supported by Cusanuswerk}\ \ and 
O. Nachtmann}
}

\date{\small\sl 
Institut  f\"ur Theoretische Physik, Universit\"at Heidelberg\\
Philosophenweg 16, D-69120 Heidelberg, Germany
}

\maketitle

%
%
%
\begin{center}
\parbox[t]{\textwidth}{\small
We consider hydrogen--like atoms in unstable levels of principal quantum number $n=2$, confined to a finite size region in a non--homogeneous electric field carrying handedness. The interplay between the internal degrees of freedom of the atoms and the external ones of their c.m.\ motion can produce P--odd contributions to the eigenenergies.
The nominal order of such shifts is $10^{-8}\,$Hz. Typically such energy shifts depend linearly on the small P--violation parameters $\deli\simeq10^{-12}$ ($i=1,2$), essentially the ratios of the P--violating mixing matrix elements of the $2S$ and $2P$ states over the Lamb shift, with $i=1$ ($i=2$) corresponding to the nuclear spin independent (dependent) term.
We show how such energy shifts can be enhanced by a factor of $\simeq 10^6$ in a resonance like way for special field configurations where a crossing of unstable levels occurs, leading to P--violating effects proportional to $\sqrt{\deli}$. Measurements of such effects can give information concerning the ``spin crisis'' of the nucleons.\vspace{1cm}\\
PACS. 11.30.Er, 21.10.Hw, 31.15.Md, 31.50.+w, 31.90.+s, 32.80.Ys.\\
Submitted to Phys.\ Lett.\ A
} 
\end{center}

\end{titlepage}

In this article we deal with parity (P)--violating effects
in atoms which are described in the standard model (SM)
of elementary particle physics by the exchange of the Z--boson between
the electrons in the shell and the quarks in the nucleus.
There is a rich literature on this subject starting from the
classical papers \cite{Bouchiat74,Moskalev76}. For a review we refer to \cite{Khriplovich91}.
Experiments on P--violating effects in atoms have been
done so far only with heavy atoms containing many electrons. Such
experiments have given information on properties of the weak
interactions and of the nuclei. The usefulness of this information
has recently been demonstrated once again \cite{PVandHERA,Wood97}.

Here we address the following question: Can one have real (complex)
eigenenergies of stable (unstable) atomic states in external
fields with a nontrivial dependence on the P--violating
parameters? If this is possible, then one could determine
the parameters of atomic P--violation by frequency
measurements, which experimentally can be pushed to very high
accuracy.

In the following we consider as an example the ordinary hydrogen atom
$\eeH$ in states with principal quantum number $n=2$. In vacuum,
and neglecting P--violation, the energy levels are $2P_{1/2}, F=0;\ 2P_{1/2},
F=1;\ 2S_{1/2}, F=0;\ 2S_{1/2}, F=1;\ 2P_{3/2}, F=1;\ 2P_{3/2},\ F=2$, with a degeneracy $2F+1$, respectively.
Here $F$ is the total angular momentum quantum number.

We denote the difference of the centers of the $2S_{1/2}$ and $2P_{1/2}$
energies, the Lamb shift, by $L$, the one of the $2P_{3/2}$ and $2P_{1/2}$
states by $\Delta$, and the hyperfine splittings by:
\bea
\label{eq1}
  \widetilde{A_1} &=& E(2S_{1/2}, F=1)-E(2S_{1/2}, F=0),\nonumber\\
  \widetilde{A_2} &=& E(2P_{1/2}, F=1)-E(2P_{1/2}, F=0),\nonumber\\
  \widetilde{A_3} &=& E(2P_{3/2}, F=2)-E(2P_{3/2}, F=1)
\eea
The values of these quantities and of the decay widths $\Gamma_S,
\Gamma_P$ of the $S$ and $P$ states are experimentally well known
\cite{Erickson77} and theoretically well understood \cite{Sapirstein90}. 
In the SM
$Z$--boson exchange induces a P--violating contribution to the Hamiltonian, $\hpv=\hpve+\hpvz$, where $\hpve (\hpvz)$ is the nuclear spin independent (dependent) contribution (for all our conventions and notations cf.\ \cite{BBN95}).
This leads to a mixing of the states $2S_{1/2}, F$ and
$2P_{1/2},F;\ 2P_{3/2},F$, with the same $F$. Numerically this
mixing is governed by two parameters $\delez$, essentially
the ratios of the relevant matrix elements of $\hpvez$
and the Lamb shift. To be precise, we define (cf.\ (3.17) of \cite{BBN95})
\be
\label{eq2}
  \deli = -\frac{G}{64\pi}\sqrt{\frac{3}{2}}
          \frac{\qwi}{m_e}\frac{1}{r_B^4 L},
          \quad(i=1,2).
\ee
Here $G$ is Fermi's constant, $r_B$ the Bohr radius and $\qwi$
are the weak charges of the proton. In the SM we have with $\theW$
the weak mixing angle
\bea
\label{eq3}
  \qwe &=& 1-4\sin^2\theW,\nonumber\\
  \qwz &=& -2(1-4\sin^2\theW)(\Del u_p-\Del d_p-\Del s_p),
\eea
where $\Delta q_p\,(q=u,d,s)$ are the quark contributions to the
total proton spin. The difference $\Delta u_p-\Delta d_p=g_A$
is the well--known axial coupling constant of neutron $\beta$--decay
and $\Delta s_p$ the spin contribution of $s$--quarks which is
subject to great current interest in connection with the so-called
``spin crisis'' of the nucleons \cite{Ashman89,Adams94}.
For a recent review cf.\ \cite{Devenish97}. Numerically the constants
$\delta_i$ are very small:
\bea
\label{eq4}
  \dele &=& -4.91\times 10^{-13},\nonumber\\
  \delz &=&  1.23\times 10^{-12}(1-g^{-1}_A\Delta s_p).
\eea
For the numerics below we set $\Delta s_p=0$.

Let us now consider the hydrogen atom in a static external
electric field $\cEv$. It has been discussed in detail
in \cite{BBN95} that as a consequence of T--invariance,
the energy levels of our atom in a {\it spatially constant} field 
(including the case of zero field, i.e.\ vacuum) 
get contributions from $\hpv$ which are at most
of order $\dele^2,\delta^2_2,\dele\delz$ and thus extremely small.

We have now investigated atoms in {\it spatially
non--homogeneous} external electric fields carrying
{\it handedness} or {\it chirality}. We give a brief account
of our results for $\eeH$ here. The detailed calculations
will be published elsewhere \cite{BGNI97}. We find that in
a field with nonzero handedness energy levels -- which are now
due to an interplay of the internal motion and the c.m.
motion of the atom -- will in general get contributions
{\it linear} in the parameters $\delta_{1,2}$. For
unstable states in special field configurations the complex
energy eigenvalues get even contributions proportional to
$\sqrt{\delta_{1,2}}$.

As a specific example we discuss the situation of a hydrogen
atom in a rectangular box with 3 segments $(\sigma=1,2,3)$ and constant
fields $\cEv^\iKsig$ in the segments
(Fig.\ 1). This situation is not quite realistic, since
Maxwell's equations require then charges on the interfaces of the
segments. But for a demonstration of the principles of P--violating energy  
shifts this does not matter and the above situation
is convenient to be analysed theoretically as a model for
spatially ``abrupt'' changes of the fields.
Spatially ``adiabatic'' changes of the fields will be
discussed elsewhere \cite{BGNII98}.

We assume that we have a field $\cE\evd$, the same in all
segments and pointing in 3--direction, and small additional fields
$\cEv'^\iKsig$ $(\sigma=1,2,3)$ which we treat as perturbations.
An atom is assumed to move in such a field configuration inside
the box. The fields in the box segments induce then Stark--mixings,
but different ones in each segment.

Through the transition conditions for the wave function at the interfaces
the internal and c.m. degrees of freedom of the atom get coupled. Then, the
quasi--stationary states in the box, i.e.\ the states with
single exponential decay, are complicated superpositions
of all $n=2$ states with various wavefunctions for the c.m.\ motion.
Now T--invariance is no longer powerful enough to exclude energy
shifts proportional to $\delta_{1,2}$. As an example we cite our
results for the following values of the box size and of the
electric fields:\\
\be
\label{eq5a}
   A_1=A_2=A_3=8\ \mathrm{nm},
\ee
\be
\label{eq5b}
   R_1^\iKe/A_1=0.45,\quad R_1^\iKz/A_1=0.55.
\ee
Here, the $A_i,\,i=1,2,3$ denote the total lengths of the the box in 1-, 2- and 3--direction respectively and the $R_1^\iKsig$ determine the widths of segments $\cS^\iKsig$ ($\sig=1,2,3$), which are defined through
\be
\label{eq5c}
   \cS^\iKsig
   :=\{\Rv=(R_1,R_2,R_3)\,|\,
                R_1^\iKvar{\sig-1}<R_1<R_1^\iKsig,\,
                0<R_2<A_2,\,0<R_3<A_3\}
\ee
($R_1^\iKn:=0,\,R_1^\iKd:=A_1$). For the electric fields we take as an example where the effect is ``large'':
\bea
\label{eq6a}
   \cE            &=& 340\,\mathrm{V/cm},\\
   {\cEv'}^\iKsig &=& \cE'\etav^\iKsig,
\eea
with 
\be
\label{eq6b}
   \cE' = 84\,\mathrm{V/cm},
\ee
\be
\label{eq6c}
   \etav^\iKe = \lk(\ba{c}1\\ 0  \\ 0\ea\rk),\quad
   \etav^\iKz = 0,\quad
   \etav^\iKd = \lk(\ba{c}0\\ 0.5\\ 0\ea\rk).
\ee 
We calculate then as observables for P--violation the differences of energies of corresponding levels in the above electric field configuration $\cE\evd+{\cEv'}^\iKsig$ and in $\cE\evd+{\cEvR'}^\iKsig$ obtained by a reflection R on the plane $x_2=A_2/2$:
\be
\label{eq8}
  R:\ \lk(\ba{c}x_1\\ x_2\\x_3\ea\rk)\longmapsto
      \lk(\ba{c}x_1\\A_2-x_2\\x_3\ea\rk).
\ee
Of course, this reflection just means a sign change in $\eta_2^\iKsig$ in (\ref{eq6c}). In this way we obtain for the lowest level which evolves continuously from the vacuum level $2S_{1/2},\,F=1,\,F_3=1$, as the confinement in the box and the fields are turned on, the following result:
\bea
\label{eq7}
   \Re (E(\cE,\{{\cEv'}^\iKsig\})-E(\cE,\{{\cEvR'}^\iKsig\})/h
   &=& \lk(\ 2.4\,\dele - 40.4\,\delz + \cO([\cE'/\cE]^3)\rk)\ {\rm kHz},
   \\
   \Im (E(\cE,\{{\cEv'}^\iKsig\})-E(\cE,\{{\cEvR'}^\iKsig\})/\hbar
   &=& \lk(-2.2\,\dele + 17.3\,\delz + \cO([\cE'/\cE]^3)\rk)\ 10^4 {\rm s^{-1}}
   \nonumber
\eea
The coefficients of the parameters $\deli$ were calculated in perturbation theory up to second order in $\cE'/\cE$.

We thus have demonstrated the existence of P--violating energy shifts
linear in $\delta_{1,2}$. But from (\ref{eq7}) we see that these shifts
are still small, of order $10^{-8}\,$Hz (cf.\ (\ref{eq4})), 
and presumably hard to
measure. We will show now that there exists the possibility
of having much bigger P--violating energy shifts proportional to
$\sqrt{\delta_{1,2}}$. To see this, consider the two states
$2S_{1/2},\,F=1,\,F_3=\pm1$ in vacuum in the absence of P--violation,
i.e.\ for $\dele=\delz=0$. These states are unstable and
have the same complex energy eigenvalue. Let us now ``turn on''
the confinement in the box, the P--violation and the zero order electric
field $\cE\evd$. It can be shown \cite{BGNI97} that T--invariance
still guarantees a twofold degeneracy of the complex eigenenergies
for the states evolving from the above ones at $\cE=0$, $\delta_{1,2}
=0$. Turning on the fields ${\cEv'}^\iKsig\ (\sigma=1,2,3)$,
we lift, in general, this degeneracy and obtain two
distinct levels. But we can search for values of
${\cEv'}^\iKsig\not=0$ where these two levels cross again
if $\dele=\delz=0$. The crossing must be in the real
{\it and} imaginary parts of the complex energy levels. The
calculation is performed using convenient versions of degenerate
perturbation theory (in the ${\cEv'}^\iKsig$--couplings of the
atom) (cf.\ \cite{BGNI97,Messiah,Bloch58}). 
Since we are dealing with unstable states
we typically get a {\it non--hermitian} 2$\times$2
matrix $\widetilde{\unH}$ (cf.\ \cite{BGNI97}), whose complex eigenvalues are the
sought energies. The structure of this matrix is again restricted
by T--invariance to be of the form
\be
\label{eq10}
  \widetilde{\unH}
  =\left(\ba{cc}
     \widetilde{H}_{++}&\widetilde{H}_{+-}\\
     \widetilde{H}_{-+}&\widetilde{H}_{--}\ea\right)
\ee
with $\widetilde{H}_{++}=\widetilde{H}_{--}$. Then the eigenvalues are
\be
\label{eq11}
  E_\pm = \widetilde{H}_{++}\pm\sqrt{\widetilde{H}_{+-}\widetilde{H}_{-+}}.
\ee
Expanding the matrix elements $\widetilde{H}_{m'm}$ in powers of $\delta_{1,2}$
we have
\bea
\label{eq12}
  \widetilde{H}_{m'm}
  &=& \widetilde{H}_{m'm}^\iKn
     +\dele\widetilde{H}_{m'm}^\iKe
     +\delz\widetilde{H}_{m'm}^\iKz
     +...,\nonumber\\
   && \qquad (m',m\in\{+1,-1\}).
\eea
For a {\it non--hermitian} matrix $\widetilde{\unH}$ we can have
\bea
\label{eq13}
  \widetilde{H}_{+-}^\iKn &=&     0,\nonumber\\
  \widetilde{H}_{+-}^\iKi &\not=& 0,\quad {\rm}\quad i=1,2,\nonumber\\
  \widetilde{H}_{-+}^\iKn &\not=& 0.
\eea
This gives
\be
\label{eq14}
  E_{\pm} = \widetilde{H}_{++}^\iKn
            \pm\sqrt{\sum^2_{i=1}\delta_i
               \widetilde{H}_{+-}^\iKi\widetilde{H}_{-+}^\iKn}
            +\cO(\dele,\delz).
\ee
We found that the conditions (\ref{eq13}) are fulfilled for instance
for a zero order field $\cE=100$ V/cm and fields ${\cEv'}^\iKsig=:\cE'\etav^\iKsig$ with
\bea
\label{eq15}
  \etav^\iKe &=& \lk(\ba{c}3.91\\ 0.0764\\ 0\ea\rk) + \cO(\cE'/\cE)\nonumber\\
  \etav^\iKz &=& 0\nonumber\\
  \etav^\iKd &=& \lk(\ba{c}3.91\\ -0.328\\ 0\ea\rk) + \cO(\cE'/\cE)
\eea
The higher order terms are under control for $\cE'<5$ V/cm, i.e.\ they
can shift the ``magic'' values of $\etav^\iKsig$ (\ref{eq15})
only slightly.

In Figs.\ 2a,\,b we show the resulting behaviour of the real and imaginary
parts of $E_{\pm}(\cEv^{'\iKsig})-E_{\pm}(\cEvR^{'\iKsig})$ as a
function of the 2nd component of the electric field in the 3rd segment, 
${\cE'_2}^\iKd=\cE'\eta_2^\iKd$, keeping all other field components fixed, i.e.\ we vary only $\eta_2^\iKd$. We see from (\ref{eq15}) that for $\cE'=5\,$V/cm,
the ``magic'' value is
\be
\label{eq16}
  {\cE'}^\iKd_{2,res}=-1.638\cdot(1+\cO(\cE'/\cE))\,\mathrm{V/cm}.
\ee
P--violation induces a sharp ``resonance'' at ${\cE'_2}^\iKd={\cE'}^\iKd_{2,res}$ where the P--violating energy shift is enhanced by six orders of magnitude. The width of the peak is proportional to $\sqrt{\deli}$ and found to be $2.6\cdot10^{-12}\,$V/cm.
Note, however, that in Figs.\ 2a,\,b the differences of the real parts of the energies of the levels
are at maximum of the order of $10^{-5}\,$Hz whereas the lifetime of the levels is calculated to be $4.13\cdot10^{-8}\,\mathrm{s}$. This corresponds to a line width $\Delta E/h = 3.86\cdot10^6\,$Hz. Thus it is certainly not an easy task to detect frequency shifts of order $10^{-5}\,$Hz in such broad lines.
But our example is only meant to provide an illustration how energy shifts proportional to $\sqrt{\delez}$ can be obtained. No effort has been made here to optimize the effect for a realistic experimental situation.

To conclude: In this paper we have demonstrated that the eigenenergies
of a hydrogen atom in a spatially inhomogeneous electric
field with nonzero chirality get in
general contributions {\it linear} in the P--violation parameters
$\delta_{1,2}$ (\ref{eq2}). This statement is valid both for stable
and unstable states and, of course, easily extended to other atoms.
To measure such P--violating contributions to eigenenergies one would
look at the {\it difference} of their values for a given
electric field configuration and its spatially reflected one.
For special electric field configurations with nonzero handedness
where two {\it unstable} levels have the same complex eigenenergy
in the absence of P--violation, we can obtain large P--violating energy shifts
proportional to $\sqrt{\delta_{1,2}}$. The comparison of the two levels at the crossing point with the levels for the
field configuration of opposite handedness should show a drastic
difference: Under a reversal of the handedness the P--violating shift gets multiplied by a
{\it phase factor} $i$. This means that the shifts in the real and
imaginary parts of the complex eigenenergies are exchanged. Finally
we note that for the case of hydrogen discussed here all our
energy shifts are much more sensitive to the parameter $\delz$
than to $\dele$ (cf.\ (\ref{eq1}--\ref{eq3})). The reason for this is the
same as given in (6.1), (6.2) of \cite{BBN95}. Thus a measurement
of our effects could lead to a determination of the
nuclear spin--dependent weak charge $\qwz$ (\ref{eq3}) which
depends on the contribution of $s$--quarks to the proton spin. In
this way parity violation in atoms could give information concerning the
``spin crisis'' of the nucleons.

\ \vspace{1cm}\\
{\bf Acknowledgements:} The authors would like to thank H. Abele, J. Kluge, H.D. Liesen, P. Overmann, W. Quint, P. Sandars, L.M. Sehgal and V. Telegdi for useful discussions and suggestions.

\newpage
\newpage
\setlength{\unitlength}{1.4mm}
\begin{picture}(100,80)
\put(35,78){\bf Figure 1}
\put( 5,15){\line(3,1){60}}
\put(35, 5){\line(3,1){60}}
\put( 5,45){\line(3,1){60}}
\put(35,35){\line(3,1){60}}
\put( 5,15){\line(3,-1){30}}
\put(65,35){\line(3,-1){30}}
\put( 5,45){\line(3,-1){30}}
\put(65,65){\line(3,-1){30}}
\put( 5,15){\line(0, 1){30}}
\put(35, 5){\line(0, 1){30}}
\put(65,35){\line(0, 1){30}}
\put(95,25){\line(0, 1){30}}
\multiput(55,11.66)(0,3){10}{\circle*{0.2}}
\multiput(75,18.33)(0,3){10}{\circle*{0.2}}
\multiput(25,21.66)(0,3){10}{\circle*{0.2}}
\multiput(45,28.33)(0,3){10}{\circle*{0.2}}
\multiput(25,21.66)(3,-1){10}{\circle*{0.2}}
\multiput(45,28.33)(3,-1){10}{\circle*{0.2}}
\multiput(25,51.66)(3,-1){10}{\circle*{0.2}}
\multiput(45,58.33)(3,-1){10}{\circle*{0.2}}
\put(12,10.66){\vector(-3, 1){10}}
\put(22, 7.33){\vector( 3,-1){10}}
\put(48, 7.33){\vector(-3,-1){10}}
\put(88,20.66){\vector( 3, 1){10}}
\put( 2,24   ){\vector( 0,-1){10}}
\put( 2,34   ){\vector( 0, 1){10}}
\put(47.5, 4.16){\vector(-3,-1){5}}
\put(57.5, 7.50){\vector( 3, 1){5}}
\put(57, 4.33){\vector(-3,-1){10}}
\put(77,11.00){\vector( 3, 1){10}}
\put(15,45){(1)}
\put(35,51.66){(2)}
\put(55,58.33){(3)}
\put(68,13){$A_1$}
\put(14, 7){$A_2$}
\put( 0,29){$A_3$}
\put(51.0, 5.33){$R_1^\iKe$}
\put(65.5, 7.16){$R_1^\iKz$}
\put(39, 8){$\eve$}
\put(29, 8){$\evz$}
\put(36,11){$\evd$}

\thicklines
\put(24,15.33){\vector(1,2){12}}
\put(50,20   ){\vector(0,1){30}}
\put(64,24.66){\vector(1,4){8}}
\put(35, 5){\vector( 3, 1){6}}
\put(35, 5){\vector(-3, 1){6}}
\put(35, 5){\vector( 0, 1){8}}
\thinlines
\put(53,49){$\vec{\cal E}$}
\end{picture}

\ \vspace{1cm}\\
\begin{description}
\item[{\sc Fig.} {\rm 1.}]
Sketch of the electric field configuration in a box divided into three segments (1)--(3). The electric field is homogeneous within every segment and varies suddenly along the 1--direction at the interfaces.
\end{description}


\newpage
\setlength{\unitlength}{0.8mm}

\begin{center}
\begin{picture}(200,95)

\put(0,0){\epsfig{file=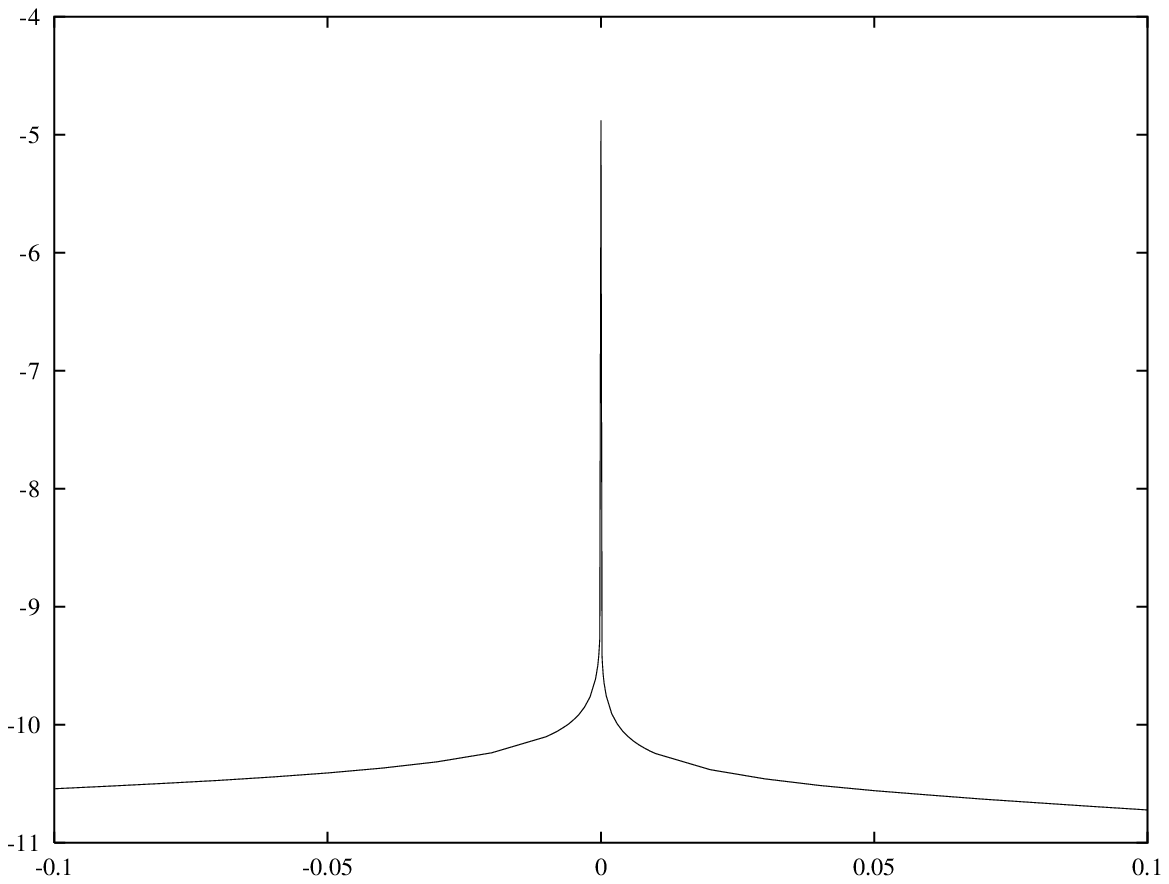,width=180\unitlength,height=100\unitlength}}


\put(75,105){\bf Figure 2a}
\put( 75,-10){$\cE_2^{'\iKd}-\cE_{2,res}^{'\iKd}$ [V/cm]}

\begin{rotate}{90}
\put(10,0){$\lg|\Re\lk(E_{\pm}(\cEv^{'\iKsig})-E_{\pm}(\cEvR^{'\iKsig})\rk)/(h\,{\mathrm Hz})|$}
\end{rotate}

\end{picture}
\ \vspace{2cm}\nopagebreak\newline
\begin{picture}(200,95)

\put(0,0){\epsfig{file=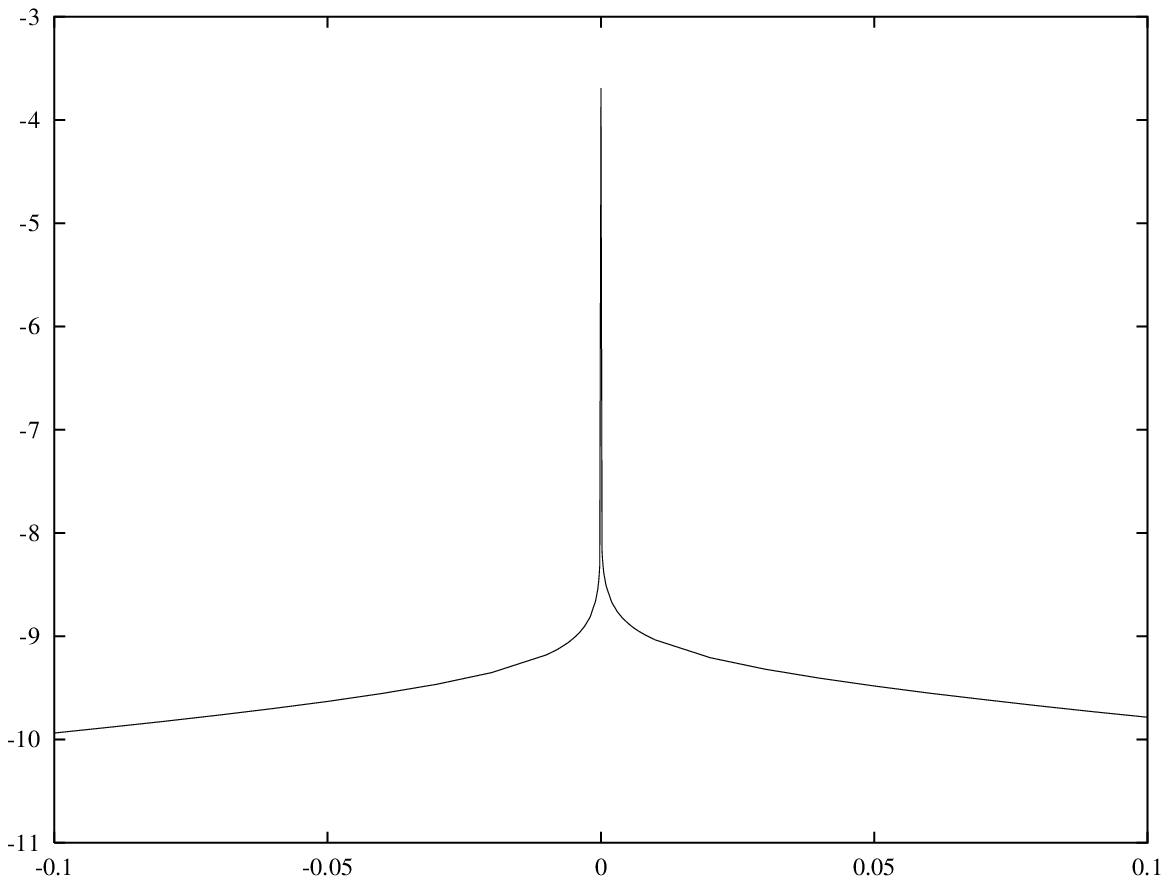,width=180\unitlength,height=100\unitlength}}


\put(75,105){\bf Figure 2b}
\put( 75,-10){$\cE_2^{'\iKd}-\cE_{2,res}^{'\iKd}$ [V/cm]}

\begin{rotate}{90}
\put(10,0){$\lg|2\Im\lk(E_{\pm}(\cEv^{'\iKsig})-E_{\pm}(\cEvR^{'\iKsig})\rk)/(\hbar\,{\mathrm s^{-1}})|$}
\end{rotate}

\end{picture}
\end{center}

\small
\ \vspace{0.5cm}\\
\begin{description}
\item[{\sc Fig.} {\rm 2.}]
The decadic logarithm of the absolute values of the real (a) and imaginary (b) parts of the P--violating energy--difference $E_{\pm}(\cEv^{'\iKsig})-E_{\pm}(\cEvR^{'\iKsig})$ (\ref{eq14}) vs.\ the deviation $\cE_2^{'\iKd}-\cE_{2,res}^{'\iKd}$ of $\cE_2^{'\iKd}$ from the resonance value (\ref{eq16}) for which the P--even splitting of levels is removed, revealing a P--violating splitting of the order of $\sqrt{\delez}$. 
\end{description}


}
\end{document}